\documentclass[final,5p,times,twocolumn]{elsarticle}

\usepackage{amssymb}
\usepackage{amsmath}
\usepackage{bm}

\journal{Physica B}

\begin{document}

\begin{frontmatter}


\title{Kondo effect in the seven-orbital Anderson model hybridized with $\Gamma_8$ conduction electrons}

\author{Takashi Hotta\corref{cor}}\ead{hotta@tmu.ac.jp}
\address{Department of Physics, Tokyo Metropolitan University,
1-1 Minami-Osawa, Hachioji, Tokyo 192-0397, Japan}
\cortext[cor]{Corresponding author}

\begin{abstract}
We clarify the two-channel Kondo effect in the seven-orbital Anderson model
hybridized with $\Gamma_8$ conduction electrons
by employing a numerical renormalization group method.
From the numerical analysis for the case with two local $f$ electrons,
corresponding to Pr$^{3+}$ or U$^{4+}$ ion,
we confirm that a residual entropy of $0.5\log 2$,
a characteristic of two-channel Kondo phenomena,
appears for the local $\Gamma_3$ non-Kramers doublet state.
For further understanding on the $\Gamma_3$ state,
the effective model is constructed on the basis of a $j$-$j$ coupling scheme.
Then, we rediscover the two-channel $s$-$d$ model concerning quadrupole
degrees of freedom.
Finally, we briefly introduce our recent result on the two-channel Kondo effect
for the case with three local $f$ electrons.
\end{abstract}

\begin{keyword}
Two-channel Kondo effect, effective model, $j$-$j$ coupling scheme, numerical renormalization group method
\end{keyword}

\end{frontmatter}


\section{Introduction}

The Kondo effect occurring in a dilute magnetic impurity system has been
understood almost completely both from theoretical and experimental viewpoints.
Then, our interests have moved onto a problem of impurity with complex degrees of freedom.
In particular, rich phenomena originating from orbital degrees of freedom have been
actively discussed for a long time.
When an impurity spin is hybridized with multichannel conduction bands,
the concept of multi-channel Kondo effect has been proposed \cite{Nozieres}.
In particular, for the case of impurity spin $1/2$ and two conduction bands,
corresponding to the overscreening situation,
it has been shown that non-Fermi liquid ground state appears.
Such non-Fermi liquid properties have been pointed out
also in a two-impurity Kondo system \cite{Jones1,Jones2}.

As for the reality of two-channel Kondo effect,
Cox has pointed out that two screening channels exist in the case of quadrupole degree of freedom
in a cubic uranium compound with non-Kramers doublet ground state \cite{Cox1,Cox2}.
In recent decades, the two-channel Kondo phenomena have been continuously and
widely investigated by many researchers at the stage of Pr compounds
\cite{review}.
We strongly believe that it is meaningful to expand the research frontier
of the two-channel Kondo physics to other rare-earth compounds.

In this paper, we discuss the two-channel Kondo effect
in the seven-orbital impurity Anderson model
hybridized with $\Gamma_8$ conduction electrons.
In order to confirm the validity of our model for the investigation of
the two-channel Kondo effect,
we consider the case with two local $f$ electrons
corresponding to Pr$^{3+}$ or $U^{4+}$ ion.
Then, we find a residual entropy of $0.5 \log 2$
as a clear signal of the two-channel Kondo effect
for the case of the non-Kramers $\Gamma_3$ doublet ground state.
By analyzing the $\Gamma_3$ state on the basis of a $j$-$j$ coupling scheme,
we obtain the two-channel $s$-$d$ model concerning
quadrupole degrees of freedom.
As an example of the development of the two-channel physics,
we briefly report our recent result on the two-channel Kondo effect
for the case with three local $f$ electrons.

\section{Analysis of Seven-Orbital Anderson Model}

The local $f$-electron Hamiltonian is given by
\begin{equation}
\label{Hloc}
\begin{split}
  H_{\rm loc} &=\sum_{m_1 \sim m_4}\sum_{\sigma,\sigma'}
  I_{m_1m_2,m_3m_4}
  f_{m_1\sigma}^{\dag}f_{m_2\sigma'}^{\dag}
  f_{m_3\sigma'}f_{m_4\sigma} +E_f n \\
 &+ \lambda \sum_{m,\sigma,m',\sigma'}
   \zeta_{m,\sigma;m',\sigma'} f_{m\sigma}^{\dag}f_{m'\sigma'}
 + \sum_{m,m',\sigma} B_{m,m'}
      f_{m \sigma}^{\dag} f_{m' \sigma},
\end{split}
\end{equation}
where $f_{m\sigma}$ is the annihilation operator
for a local $f$ electron with spin $\sigma$ and $z$-component $m$
of angular momentum $\ell=3$,
$\sigma=+1$ ($-1$) for up (down) spin,
$I$ indicates Coulomb interactions,
$E_f$ is an $f$-electron level,
$n$ denotes the local $f$-electron number,
$\lambda$ is the spin-orbit coupling,
and $B_{m,m'}$ indicates crystalline electric field (CEF) potentials.

The Coulomb interaction $I$ is expressed as
\begin{equation}
I_{m_1m_2,m_3m_4} = \sum_{k=0}^{6} F^k c_k(m_1,m_4)c_k(m_3,m_2),
\end{equation}
where $F^k$ indicates the Slater-Condon parameter and
$c_k$ is the Gaunt coefficient \cite{Slater}.
The sum is limited by the Wigner-Eckart theorem to
$k=0$, $2$, $4$, and $6$.
Although the Slater-Condon parameters should be determined
for the material from the experimental results,
here we set the ratio as
\begin{equation}
 \label{SlaterCondon}
  F^0=10U,~F^2=5U,~F^4=3U, F^6=U,
\end{equation}
where $U$ is the Hund rule interaction among $f$ orbitals.
In the spin-orbit coupling term,
each matrix element of $\zeta$ is given by
\begin{equation}
\begin{split}
\zeta_{m,\sigma;m,\sigma}&=m\sigma/2,\\
\zeta_{m+\sigma,-\sigma;m,\sigma}&=\sqrt{\ell(\ell+1)-m(m+\sigma)}/2,
\end{split}
\end{equation}
and zero for other cases.
The CEF potentials for $f$ electrons from the ligand ions are given
in the table of Hutchings for the angular momentum $\ell=3$ \cite{Hutchings}.
For $O_{\rm h}$ symmetry, $B_{m,m'}$ is expressed by two CEF parameters,
$B_4^0$ and $B_6^0$, as
\begin{equation}
\begin{split}
    B_{3,3}&=B_{-3,-3}=180B_4^0+180B_6^0, \\
    B_{2,2}&=B_{-2,-2}=-420B_4^0-1080B_6^0, \\
    B_{1,1}&=B_{-1,-1}=60B_4^0+2700B_6^0, \\
    B_{0,0}&=360B_4^0-3600B_6^0, \\
    B_{3,-1}&=B_{-3,1}=60\sqrt{15}(B_4^0-21B_6^0),\\
    B_{2,-2}&=300B_4^0+7560B_6^0.
\end{split}
\end{equation}
Note the relation $B_{m,m'}=B_{m',m}$.

Now, we consider the case of $n=2$ by appropriately adjusting the value of $E_f$.
As $U$ denotes the magnitude of the Hund rule interaction among $f$ orbitals,
it is reasonable to set $U=1$ eV.
The magnitude of $\lambda$ varies between 0.077 and 0.36 eV
depending on the type of lanthanide ions.
For a Pr$^{3+}$ ion, $\lambda$ is $720-730$ cm$^{-1}$ \cite{Carnall}.
Thus, we set $\lambda=0.09$ eV.
In Fig.~1, we depict the ground-state phase diagram of $H_{\rm loc}$ for
$n=2$ on the plane of $B_4^0$ and $B_6^0$.
For negative $B_6^{0}$, we find $\Gamma_5$ and $\Gamma_1$ states
depending on the values of $B_4^0$,
while for positive $B_6^0$, we observe $\Gamma_3$ state at a region 
including $B_4^0=0$, sandwiched by the $\Gamma_5$ and $\Gamma_1$ states.

\begin{figure}[t]
\centering
\includegraphics[width=0.98\linewidth]{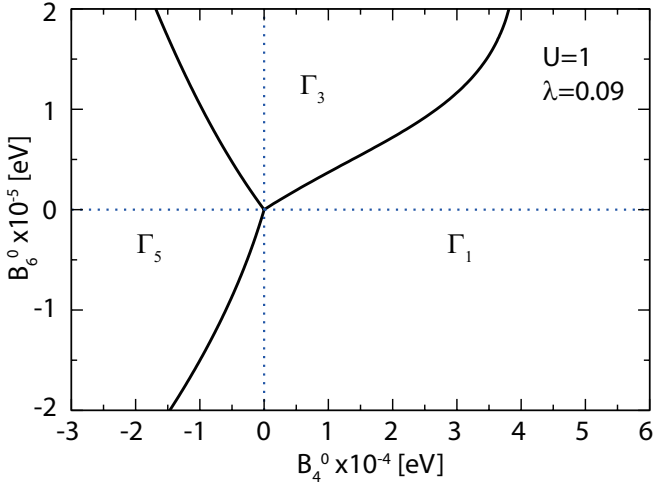}
\caption{
Ground-state phase diagram of $H_{\rm loc}$
on the $(B_4^0, B_6^0)$ plane for $n=2$.
}
\end{figure}

Next, we include the hybridization with $\Gamma_8$ conduction electron bands.
For the purpose, we transform the $f$-electron basis
in $H_{\rm loc}$ from $(m, \sigma)$ to $(j, \tau, \sigma)$,
where $j$ indicates the total angular momentum of one $f$-electron state,
$\tau$ denotes the irreducible representation of $O_{\rm h}$ point group,
and $\sigma$ in $(j, \tau, \sigma)$ indicates the pseudo-spin up ($\uparrow$) and
down ($\downarrow$) to distinguish the Kramers degenerate state.
Note that here we use the same $\sigma$ both for real and pseudo spins.
For $j=7/2$ octet, we have two doublets ($\Gamma_6$ and $\Gamma_7$)
and one quartet ($\Gamma_8$), while for $j=5/2$ sextet,
we obtain one doublet ($\Gamma_7$) and one quartet ($\Gamma_8$).

Then, the local Hamiltonian is given by
\begin{equation}
\label{Hloc2}
\begin{split}
  {\tilde H}_{\rm loc} & \!=\!
 \sum_{j, \tau, \sigma} ({\tilde \lambda}_j  + {\tilde B}_{j,\mu} + E_f ) f_{j \tau \sigma}^{\dag} f_{j \tau \sigma}\\
 &\!+\! \sum_{j_1 \! \sim \! j_4} \! \sum_{\tau_1 \! \sim \! \tau_4} \!
  \sum_{\sigma_1 \! \sim \! \sigma_4} \!
  {\tilde I}^{j_1 j_2, j_3 j_4}_{\tau_1 \sigma_1 \tau_2 \sigma_2, \tau_3 \sigma_3 \tau_4 \sigma_4}
  f_{j_1 \tau_1 \sigma_1}^{\dag} \! f_{j_2 \tau_2 \sigma_2}^{\dag} \!
  f_{j_3 \tau_3 \sigma_3} \! f_{j_4 \tau_4 \sigma_4},
\end{split}
\end{equation}
where we set $j=a$ ($b$) for $j=5/2$ ($7/2$),
${\tilde \lambda}_a=-2\lambda$,
${\tilde \lambda}_b=(3/2)\lambda$,
${\tilde B}_{j,\tau}$ denotes the CEF potential energy,
$f_{j \tau \sigma}$ indicates the annihilation operator of $f$ electron in the bases of $(j, \tau, \sigma)$,
and ${\tilde I}$ denotes the Coulomb interactions between $f$ electrons.

Here we assume the hybridization between
$\Gamma_8$ conduction electrons and $\Gamma_8$ quartet of $j=5/2$,
since the $j=5/2$ states should be mainly occupied for the case of $n<7$.
Then, the seven-orbital Anderson model is expressed as
\begin{equation}
  H = \sum_{\bm{k},\tau,\sigma} \varepsilon_{\bm{k}}
    c_{\bm{k}\tau\sigma}^{\dag} c_{\bm{k}\tau\sigma}
   + \sum_{\bm{k},\tau,\sigma} V(c_{\bm{k}\tau\sigma}^{\dag} f_{a\tau\sigma}+{\rm h.c.})
   + \tilde{H}_{\rm loc},
\end{equation}
where $\varepsilon_{\bm{k}}$ is the dispersion of conduction electron with wave vector $\bm{k}$,
$c_{\bm{k}\tau\sigma}$ is the annihilation operator of a $\Gamma_8$ conduction electron,
$\tau$ (=$\alpha$ and $\beta$) distinguishes the $\Gamma_8$ quartet,
and $V$ is the hybridization between conduction and localized electrons.

In order to diagonalize the impurity Anderson model, we employ
a numerical renormalization group (NRG) method \cite{NRG1,NRG2},
in which we logarithmically discretize the momentum space so as
to efficiently include the conduction electrons near the Fermi energy.
The conduction electron states are characterized by ``shells'' labeled by $N$,
and the shell of $N=0$ denotes an impurity site described by the local Hamiltonian.
Then, after some algebraic calculations,
the Hamiltonian is transformed into the recursive form
\begin{eqnarray}
  H_{N+1} = \sqrt{\Lambda} H_N + t_N \sum_{\tau,\sigma}
  (c_{N \tau \sigma}^{\dag}c_{N+1 \tau \sigma}+c_{N+1 \tau \sigma}^{\dag}c_{N \tau \sigma}),
\end{eqnarray}
where $\Lambda$ is a parameter used for logarithmic discretization,
$c_{N \tau \sigma}$ denotes the annihilation operator of the conduction electron
in the $N$-shell, and $t_N$ indicates the ``hopping'' of the electron between
$N$- and $(N+1)$-shells, expressed by
\begin{eqnarray}
  t_N=\frac{(1+\Lambda^{-1})(1-\Lambda^{-N-1})}
  {2\sqrt{(1-\Lambda^{-2N-1})(1-\Lambda^{-2N-3})}}.
\end{eqnarray}
The initial term $H_0$ is given by
\begin{equation}
  H_0=\Lambda^{-1/2} [H_{\rm loc}
  +\sum_{\tau,\sigma} V(c_{0\tau \sigma}^{\dag} f_{a\tau \sigma}
  +f_{a\tau \sigma}^{\dag}c_{0\tau \sigma})].
\end{equation}

For the calculation of thermodynamic quantities,
the free energy $F$ for the local $f$ electron is evaluated in each step as
\begin{eqnarray}
F = -T (\ln {\rm Tr} e^{-H_N/T} - \ln {\rm Tr} e^{-H_N^0/T}),
\end{eqnarray}
where a temperature $T$ is defined as $T=\Lambda^{-(N-1)/2}$
in the NRG calculation and $H_N^0$ denotes the Hamiltonian
without the impurity and hybridization terms.
Then, the entropy $S_{\rm imp}$ is obtained by
$S_{\rm imp}=-\partial F/\partial T$
and the specific heat $C_{\rm imp}$ is evaluated by
$C_{\rm imp}=-T\partial^2 F/\partial T^2$.
In the NRG calculation, $M$ low-energy states are kept
for each renormalization step.
In this paper, we set $\Lambda=5$ and $M=2,500$.

\begin{figure}[t]
\centering
\includegraphics[width=0.98\linewidth]{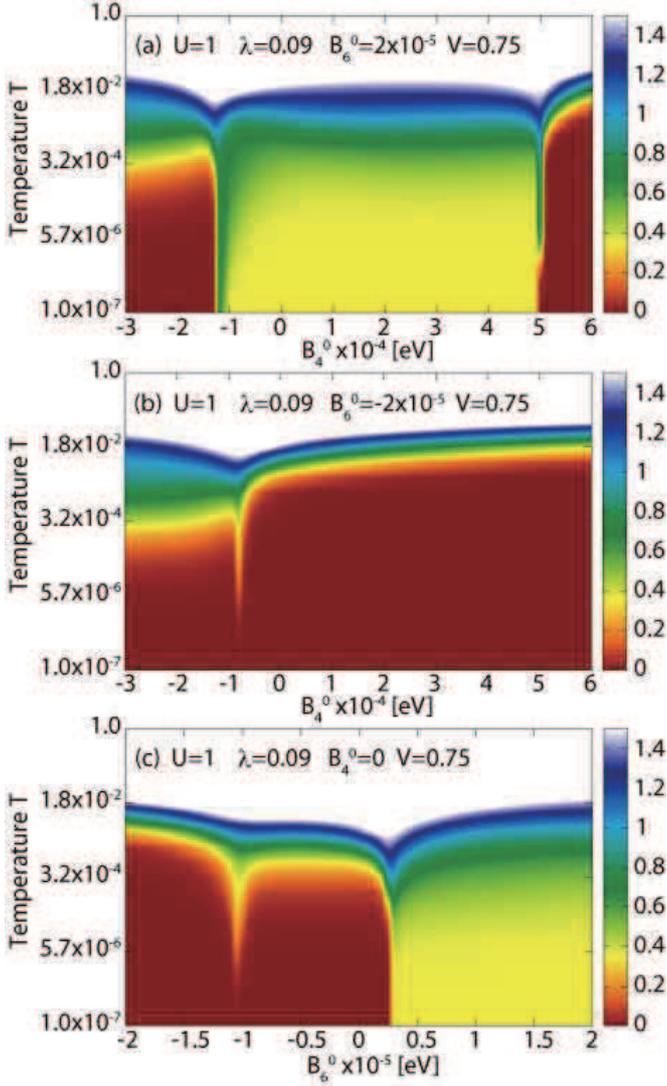}
\caption{
Color contour maps of the entropies for $n=2$ on
the plane of 
(a) ($B_4^0,  T)$ for $B_6^0=2 \times 10^{-5}$,
(b) ($B_4^0,  T)$ for $B_6^0=-2 \times 10^{-5}$,
and
(c) ($B_6^0,  T)$ for $B_4^0=0$.
Note that $T$ is given in a logarithmic scale.
}
\end{figure}

In Fig.~2(a), we show the contour color map of the entropy
on the plane of $B_4^0$ and $T$ for $B_6^0=2 \times 10^{-5}$ and $V=0.75$.
For the visualization of the behavior of entropy, the color of the entropy
is defined between $0$ and $1.5$, as shown in the right color bar.
For $B_4^0 \le -10^{-4}$, corresponding to the $\Gamma_5$ region in Fig.~1,
when a temperature is decreased, we find a short plateau of the entropy $\log 2$ (green region),
but the entropy is eventually released at low temperatures,
while for $B_4^0 \ge 5 \times 10^{-4}$, corresponding to
the $\Gamma_1$ region in Fig.~1, the entropy is promptly released at low temperatures,
without entering the plateau of the entropy $\log 2$.
However, in the region of $-10^{-4} \le B_4^0  \le 5 \times 10^{-4}$,
we widely observe a entropy $0.5\log 2$ (yellow region),
a characteristic of the two-channel Kondo effect.
We emphasize that the yellow region overlaps with the $\Gamma_3$ one in Fig.~1,
although it has a small overlap with the $\Gamma_1$ region.
Thus, we conclude that the two-channel Kondo effect widely occurs
in the $\Gamma_3$ non-Kramers doublet ground state,
as was pointed out by Cox.

In Fig.~2 (b),  we show the result for $B_6^0=-2 \times 10^{-5}$ and $V=0.75$.
For $B_4^0$ values corresponding to the $\Gamma_5$ and $\Gamma_1$ regions
in Fig.~1, in comparison with Fig.~2(a),
we find similar behavior of the temperature dependence of the entropy in each region.
In sharp contrast to Fig.~2(a), we do not find the wide yellow region in Fig.~2(b),
since for negative $B_6^0$, the $\Gamma_3$ ground state does not appear.
However, we observe a very narrow yellow region near $B_4^0=-10^{-4}$,
suggesting the entropy $0.5\log 2$ of the non-Fermi liquid behavior,
which is known to appear at a boundary point between Kondo and non-Kondo phases
corresponding to the different fixed points
in a two-orbital Anderson model \cite{Fabrizio1,Fabrizio2,Mitchell}.
In the present case, we obtain the standard Kondo effect in the $\Gamma_5$ state,
while the singlet state appears in the $\Gamma_1$ state.
Thus, near the boundary between the $\Gamma_5$ (Kondo) and $\Gamma_1$ (non-Kondo)
states, we expect the non-Fermi liquid behavior just at a critical point.

In Fig.~2(c), in order to reconfirm the above results,
we show the contour color map of the entropy
on the plane of $B_6^0$ and $T$ for $B_4^0=0$ and $V=0.75$.
As expected from Figs.~2(a) and 2(b), we find the wide yellow region
for $B_6^0 \ge 3 \times 10^{-6}$,
almost corresponding to the $\Gamma_3$ region in Fig.~1.
For negative $B_6^0$, since there is no $\Gamma_3$ region in Fig.~1,
we do not observe the wide yellow region, but again, we find the narrow
sharp yellow region near $B_6^0=-10^{-5}$.
From these results, we deduce that the curve for the critical points of
the non-Fermi liquid state runs in the $\Gamma_1$ state
along the boundary between the $\Gamma_5$ and $\Gamma_1$ regions.
It is worth while to investigate how the curve for the critical points merges
to the wide non-Fermi liquid region in the $\Gamma_3$ phase,
but this point will be discussed elsewhere in the future.

\begin{figure}[t]
\centering
\includegraphics[width=0.98\linewidth]{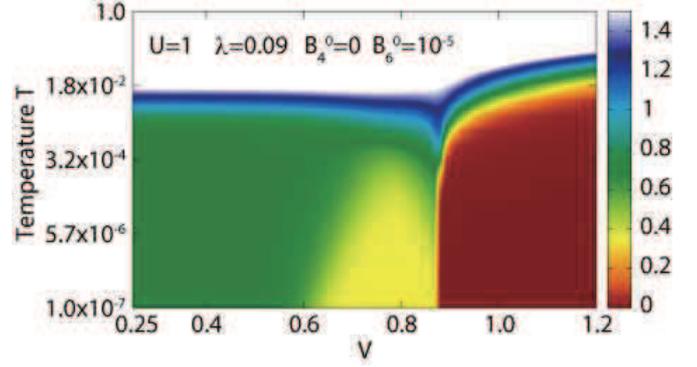}
\caption{Color contour map of entropy on the $(V, T)$ plane
for $B_4^0=0$ and $B_6^0=10^{-5}$.}
\end{figure}

Finally, let us consider the $V$ dependence of the entropy.
In Fig.~3, we show the contour map of entropy on the $(V,T)$ plane
for $B_4^0=0$ and $B_6^0=10^{-5}$ with the $\Gamma_3$ local ground state.
We emphasize  that the $0.5\log 2$ entropy does not appear
only at a certain value of $V$,
but it can be observed in the wide region of $V$
as $0.6 < V < 0.9$ in the present temperature range.
This behavior is different from that in the non-Fermi liquid state
due to the competition between CEF and Kondo-Yosida singlets
for $f^2$ systems \cite{Miyake1,Miyake2,Miyake3}.
We also remark that the two-channel Kondo effect appears for relatively
large values of $V$ in the present energy scale of $U=D=1$ eV.

\section{Analysis of Effective Model}

From the NRG calculation results on the seven-orbital Anderson model,
we believe that the two-channel Kondo effect is confirmed to occur
for the case of $n=2$ in the local $\Gamma_3$ ground state.
We also have found the non-Fermi liquid behavior just at the critical point
between $\Gamma_5$ and $\Gamma_1$ states.

However, it is difficult to describe the electronic state of
the $\Gamma_3$ non-Kramers doublet from a microscopic viewpoint,
since all the $f$ orbitals are included in the present calculations.
In order to clarify this point and visualize the $\Gamma_3$ state,
we construct the effective Hamiltonian including only $j=5/2$ states
by exploiting a $j$-$j$ coupling scheme \cite{Hotta1,Hotta2,Hattori}.

The model is given by the sum of effective Coulomb interaction and
CEF potential terms as
\begin{equation}
\label{Heff}
\begin{split}
  H_{\rm eff} & = \sum_{\tau,\sigma} ({\tilde B}_{a,\tau} + {\tilde E}_f)
  f_{a \tau \sigma}^{\dag}f_{a \tau \sigma} \\
 &+\sum_{\tau_1 \! \sim \! \tau_4} \! \sum_{\sigma_1 \! \sim \! \sigma_4}
  {\tilde I}^{aa,aa}_{\tau_1 \sigma_1 \tau_2 \sigma_2,\tau_3 \sigma_3 \tau_4 \sigma_4}
  f_{a \tau_1 \sigma_1}^{\dag} f_{a \tau_2 \sigma_2}^{\dag}
  f_{a \tau_3 \sigma_3}f_{j_4 \tau_4 \sigma_4} 
\end{split}
\end{equation}
where ${\tilde B}_{a,\tau}$ is the CEF potential for one $f$ electron in the $j=5/2$ state,
${\tilde E}_f$ denotes the $f$-electron level to adjust the local $f$-electron number,
and ${\tilde I}^{aa,aa}$ denotes the effective interaction between $f$ electrons in the $j=5/2$ states.

The CEF potential is given by
\begin{equation}
\label{CEFpotjj}
{\tilde B}_{a,\alpha}={\tilde B}_{a,\beta}=\frac{120 \cdot 11}{7}B_4^0,~
{\tilde B}_{a,\gamma}=-\frac{240 \cdot 11}{7} B_4^0,
\end{equation}
where $B_4^0$ is the same as that in Eq.~(\ref{Hloc}).
Note that $\tau=\alpha$ and $\beta$ denote $\Gamma_{8}$ states,
while $\tau=\gamma$ indicates $\Gamma_7$ state.
The factor $11/7$ is obtained from the discussion on the Stevens factor \cite{Hotta2},
but the value is in common between the $LS$ and $j$-$j$ coupling schemes,
since these two pictures provide the same results for one $f$-electron case.
Thus, this term should be always given in the present form.
We also note that the 6th-order CEF terms do not appear in the CEF potentials
in the $j$-$j$ coupling scheme,
since the maximum change in the $z$-component of total angular momentum
is equal to five in the $j=5/2$ sector.
Namely, when we use the bases of $j$,
the effect of $B_6^0$ terms appears only through the $j=7/2$ states.

In order to explain the prescription to obtain the effective Coulomb interaction ${\tilde I}^{aa,aa}$,
we separate it into two parts as
\begin{equation}
 \label{effint}
{\tilde I}^{aa,aa}_{\tau_1 \sigma_1 \tau_2 \sigma_2,\tau_3 \sigma_3 \tau_4 \sigma_4}
={\tilde I}^{(0)}_{\tau_1 \sigma_1 \tau_2 \sigma_2,\tau_3 \sigma_3 \tau_4 \sigma_4}
+{\tilde I}^{(1)}_{\tau_1 \sigma_1 \tau_2 \sigma_2,\tau_3 \sigma_3 \tau_4 \sigma_4},
\end{equation}
where ${\tilde I}^{(0)}$ denotes the Coulomb interactions among $f$ electrons
in the $j=5/2$ states in the limit of $\lambda=\infty$,
whereas ${\tilde I}^{(1)}$ indicates the correction term due to the effect of finite value
of $\lambda$.

Concerning the Coulomb interactions ${\tilde I}^{(0)}$,
here we briefly explain the way to derive them.
The matrix elements of ${\tilde I}^{(0)}$ are calculated by the Coulomb
integrals with the use of the wave functions of $j=5/2$ states.
Such Coulomb integrals are expressed by three Racah parameters,
defined as \cite{Hotta1}
\begin{eqnarray}
  \label{Racahjj}
  \begin{array}{l}
   \displaystyle
    E_0 = F^0-\frac{80}{1225}F^2-\frac{12}{441}F^4,\\
   \displaystyle
    E_1 = \frac{120}{1225}F^2+\frac{18}{441}F^4, \\
   \displaystyle
    E_2 = \frac{12}{1225}F^2-\frac{1}{441}F^4.
  \end{array}
\end{eqnarray}
Again we should note that the effect of the 6th-order Slater-Condon parameter
$F^6$ is not included in the Coulomb integrals evaluated from the $j=5/2$ states.
The explicit forms of ${\tilde I}^{(0)}$ with the use of $E_0$, $E_1$, and $E_2$
are found in the Appendix.

Next we consider the term ${\tilde I}^{(1)}$, which plays important role
to stabilize the $\Gamma_3$ state.
As mentioned above, the effect of the 6th-order CEF potential $B_6^0$ is not
included in the CEF potential term Eq.~(\ref{CEFpotjj}).
In order to include  effectively the 6th-order CEF potential terms, it is necessary
to consider the two-electron potentials, leading to the effective interaction
${\tilde I}^{(1)}$.
A straightforward way to include the effect of $B_6^0$
terms into the $j=5/2$ states
is to apply the perturbation theory in terns of $1/\lambda$
to consider effectively the contribution from the $j=7/2$ states \cite{Hotta2}.
The calculations are tedious, but it is possible to obtain systematically
the matrix elements of the effective interaction.
It is also possible to obtain such effective interactions numerically \cite{Hattori}.
In the present paper, we propose another complementary way
to obtain the analytic forms of the matrix elements of ${\tilde I}^{(1)}$
in order to promote our understanding on the $f$-electron state.

For the purpose, we consider the effective model of $H_{\rm loc}$
which reproduces well the low-energy states,
namely, the CEF states in the multiplet characterized by
the total angular momentum $J$.
Such an effective model is known as the Stevens Hamiltonian,
expressed by Stevens' operator equivalent as
\begin{equation}
  \label{StevensH}
  H_{\rm S} \!=\! B_4^0(n,J)({\hat O}_4^0+5{\hat O}_4^4)
  \!+\! B_6^0(n,J)({\hat O}_6^0-21{\hat O}_6^4),
\end{equation}
where $B_p^q(n,J)$ and ${\hat O}_p^q$ denote, respectively,
the CEF parameter and the Stevens' operator equivalent for $n$ and $J$.
The matrix elements of ${\hat O}_p^q$ for any value of $J$ have
been already tabulated by Hutchings.
Here we explicitly show the values of $n$ and $J$ in the parentheses
of the CEF parameter, since it is necessary to distinguish them from
$B_4^0$ and $B_6^0$ for $J=\ell=3$ in Eq.~(\ref{Hloc}).

Note that $H_{\rm S}$ is the effective Hamiltonian for
the multiplet specified by $J$ for $any$ values of $U$ and $\lambda$,
as long as they are sufficiently larger than the typical size of
the CEF potential energy.
For the case of $n=2$, the ground-state multiplet is characterized by $J=4$
with nine-fold degeneracy.
This degeneracy is lifted by the CEF potential into $\Gamma_1$ singlet,
$\Gamma_3$ doublet, $\Gamma_4$ triplet, and $\Gamma_5$ triplet.
These results do not depend on the values of $U$ and $\lambda$,
except for the values of the eigenenergies.

Our idea to derive the effective interaction is as follows.
We express the state of $|J, J_z \rangle$ for $n=2$ by the linear combinations
of the two-electron state $f^{\dag}_{a \tau_1 \sigma_1} f^{\dag}_{a \tau_2 \sigma_2} |0 \rangle$,
where $J_z$ denotes the $z$-component of $J$ and $|0 \rangle$ indicates the vacuum.
Since the non-zero matrix elements are obtained from the evaluation of
$\langle J, J'_z | H_S | J,J_z \rangle$, it is possible to derive the effective interaction 
among two-electron states.

Here we should pay due attention to the treatment of $B_4^0$ term in $H_S$.
As easily understood from Eqs.~(\ref{Heff}) and (\ref{CEFpotjj}),
we already include the $B_4^0$ term in the one-electron potential,
which, of course, induces the potentials for two-electron states.
Thus, if we also include the $B_4^0$ term into the potential for the
two-electron states through the evaluation of $H_S$,
such $B_4^0$ effects are doubly counted.
We note that Eq.~(\ref{CEFpotjj}) correctly provides the potentials
acting on the two-electron states. 
In order to avoid such double counting, we suppress the $B_4^0$ term
for the derivation of ${\tilde I}^{(1)}$ from $H_S$.

Then, we obtain ${\tilde I}^{(1)}$ by evaluating
$B_6^0(n,J)({\hat O}_6^0-21{\hat O}_6^4)$
with the use of the two-electron state
$f^{\dag}_{a \tau_1 \sigma_1} f^{\dag}_{a \tau_2 \sigma_2} |0 \rangle$.
We show the explicit forms of ${\tilde I}^{(1)}$ in the Appendix,
in which ${\tilde I}^{(1)}$ is expressed by $B_6$, defined as
\begin{equation}
   \label{b6}
   B_6=B_6^0(2,4) =R_{6}(2,4) B_6^0.
\end{equation}
Here $R_6(n,J)=\gamma^{(n)}_{J}/\gamma_{\ell}$
with the Stevens factor  $\gamma^{(n)}_{J}$ and
$\gamma_{\ell}=-4/(9\cdot13\cdot33)$ \cite{Stevens}.

\begin{figure}[t]
\centering
\includegraphics[width=0.98\linewidth]{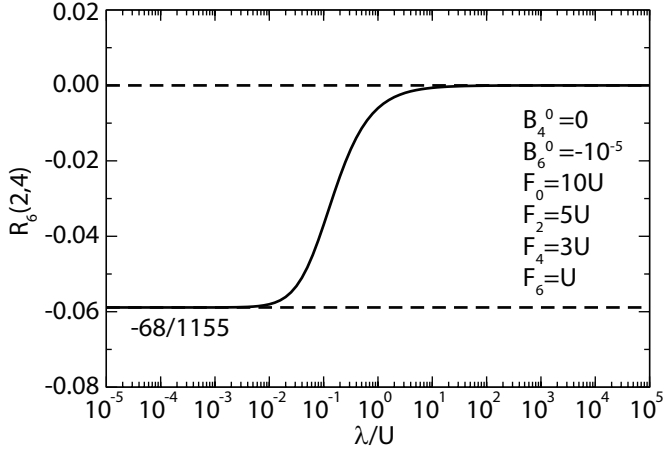}
\caption{
$R_6(2,4)$ versus $\lambda/U$ for the case of Eq.~(\ref{SlaterCondon})
with the CEF parameters of $B_4^0=0$, and $B_6^0=-10^{-5}$.
}
\end{figure}

As for the value of $R_6(2,4)$, we obtain $R_6(2,4)=-68/1155$
in the limit of $U=\infty$ (the $LS$ coupling scheme).
On the other hand, in the limit of $\lambda=\infty$ (the $j$-$j$ coupling scheme),
we find $R_6(2,4)=0$, which is quite natural,
since the $B_6^0$ terms do not appear for the $j=5/2$ states.
For finite values of $U$ and $\lambda$, we do not know the analytic value
of $R_6(2,4)$, but we can obtain it numerically.
The curve of $R_6(2,4)$ versus $\lambda/U$ is depicted in Fig.~4.
We observe that the value of $R_6(2,4)$ changes smoothly from $-68/1155$
at $\lambda/U \ll 1$ to $0$ at $\lambda/U \gg 1$.
Here we point out that in actual materials,
$\lambda/U$ is in the order of $0.1$ in the transition region
from the value in the $LS$ coupling scheme to that in the $j$-$j$ coupling scheme.
For $\lambda=0.09$ and $U=1$, we find $R_6(2,4)=-0.0388$.

Now the effective local model $H_{\rm eff}$ is ready.
In Fig.~5(a), we show the ground-state phase diagram of $H_{\rm eff}$
on the $(B_4^0, B_6^0)$ plane for the same parameters as in Fig.~1.
The basic structure of the appearance of the phases
is the same as that in Fig.~1.
Namely, for negative $B_6^{0}$, $\Gamma_5$ and $\Gamma_1$ states
are found for $B_4^0<0$ and $B_4^0>0$, respectively,
whereas for positive $B_6^0$, the $\Gamma_3$ state appears 
between $\Gamma_5$ and $\Gamma_1$ states.
The phase boundary curves are found to be deviated slightly from those in Fig.~1,
but we conclude that the local phase diagram of $H_{\rm eff}$ is essentially
the same as that of $H_{\rm loc}$ for small $B_4^0$ and $B_6^0$.

\begin{figure}[t]
\centering
\includegraphics[width=0.98\linewidth]{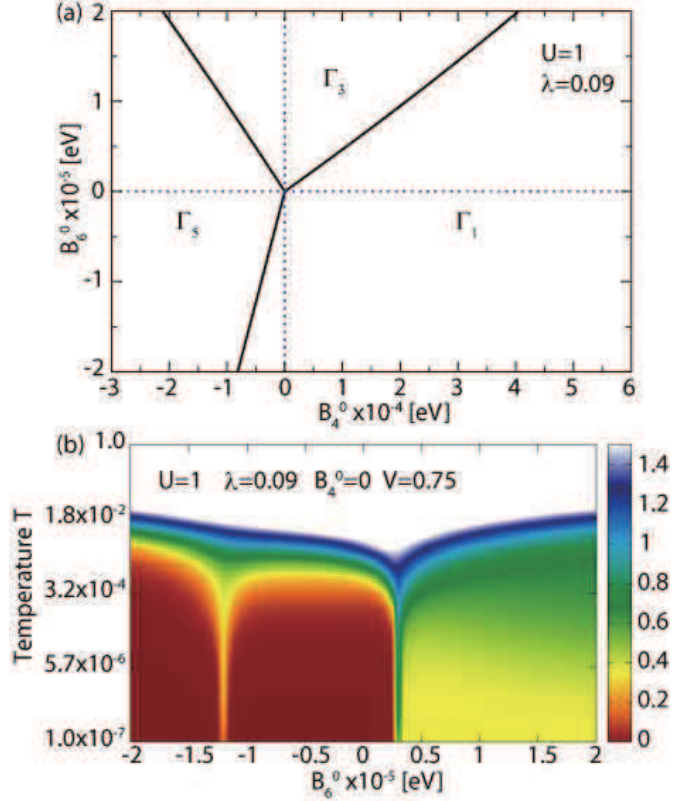}
\caption{
(a) Ground-state phase diagram of the effective local Hamiltonian
on the $(B_4^0, B_6^0)$ plane for $n=2$.
All parameters are taken as the same as those in Fig.~1.
(b) Color contour map of the entropy of the effective model
for $n=2$ on the ($B_6^0,  T)$ plane for $B_4^0=0$.
}
\end{figure}

Let us now move on to the NRG result of the effective Anderson model, given by
\begin{equation}
  H = \sum_{\bm{k},\tau,\sigma} \varepsilon_{\bm{k}}
    c_{\bm{k} \tau \sigma}^{\dag} c_{\bm{k} \tau \sigma}
   + \sum_{\bm{k},\tau, \sigma} V(c_{\bm{k} \tau \sigma}^{\dag} f_{a \tau \sigma}+{\rm h.c.})
   + H_{\rm eff}.
\end{equation}
In Fig.~5(b), we show the contour color map of the entropy of the above model
on the plane of $B_6^0$ and $T$ for $B_4^0=0$ and $V=0.75$.
We obtain essentially the same results as found in Fig.~2(c).
Namely, the wide yellow region is found for $B_6^0 \ge 3 \times 10^{-6}$,
which seems to correspond to the $\Gamma_3$ region in Fig.~5(a).
In comparison with Fig.~2(c), we point out that the green color seems
to be darker, suggesting that the regions of the plateau of $\log 2$ becomes
wider than those in Fig.~2(c).
For negative $B_6^0$, as we have found in Fig.~2(c),
we do not observe the wide yellow region,
but the narrow sharp yellow region is found near $B_6^0=-10^{-5}$.
In comparison with Fig.~2(c), the plateau of $0.5 \log 2$ is found
even at lower temperatures, clearly suggesting the existence of
the critical point between the $\Gamma_5$ and $\Gamma_1$ regions.
Note that we can also find the same behavior in Fig.~2(c), if we change
more precisely the values of $B_6^0$ near $B_6^0=-10^{-5}$.

Since the results on the effective Hamiltonian have been essentially
the same as those on the original seven-orbital model,
we believe that it is allowed to analyze the $\Gamma_3$ state
on the basis of the $j$-$j$ coupling scheme with the effective interactions.
Then, after some algebraic calculations, we find that the ground-state
$\Gamma_3$ states are expressed as
\begin{equation}
\begin{split}
|\Gamma_{3\alpha} \rangle &= \sqrt{\frac{16}{21}} |S_{78\alpha}\rangle
+\sqrt{\frac{5}{21}} |S^{(1)}_{8} \rangle,\\
|\Gamma_{3\beta} \rangle &= \sqrt{\frac{16}{21}} |S_{78\beta}\rangle
+\sqrt{\frac{5}{21}} |S^{(2)}_{8} \rangle,
\end{split}
\end{equation}
where $|S_{78 \alpha}\rangle$ and $|S_{78 \beta}\rangle$ denote
the singlet states between $\Gamma_7$ and $\Gamma_{8}$ states,
as schematically shown in Fig.~6,
while $|S^{(1)}_{8} \rangle$ and $|S^{(2)}_{8} \rangle$ denote
the singlet states in the $\Gamma_{8}$ states.
We obtain $|S_{78 \alpha}\rangle$ and $|S_{78 \beta}\rangle$,
respectively. as
\begin{equation}
\begin{split}
|S_{78 \alpha}\rangle &= \frac{1}{\sqrt{2}}
\left( f^{\dag}_{a \gamma \uparrow} f^{\dag}_{a \alpha \downarrow}
-f^{\dag}_{a \gamma \downarrow} f^{\dag}_{a \alpha \uparrow}
\right) |0\rangle,\\
|S_{78 \beta}\rangle &= \frac{1}{\sqrt{2}}
\left( f^{\dag}_{a \gamma \uparrow} f^{\dag}_{a \beta \downarrow}
-f^{\dag}_{a \gamma \downarrow} f^{\dag}_{a \beta \uparrow}
\right) |0\rangle,
\end{split}
\end{equation}
while $|S^{(1)}_{8} \rangle$ and $|S^{(2)}_{8} \rangle$ are,
respectively, given by
\begin{equation}
\begin{split}
|S^{(1)}_{8} \rangle &= \frac{1}{\sqrt{2}}
\left(f^{\dag}_{a \beta \uparrow} f^{\dag}_{a \beta \downarrow}
- f^{\dag}_{a \alpha \uparrow} f^{\dag}_{a \alpha \downarrow}
\right) |0\rangle,\\
|S^{(2)}_{8} \rangle &= \frac{1}{\sqrt{2}}
\left( f^{\dag}_{a \alpha \uparrow} f^{\dag}_{a \beta \downarrow}
-f^{\dag}_{a \alpha \downarrow} f^{\dag}_{a \beta \uparrow}
\right) |0\rangle.
\end{split}
\end{equation}

\begin{figure}[t]
\centering
\includegraphics[width=1.0\linewidth]{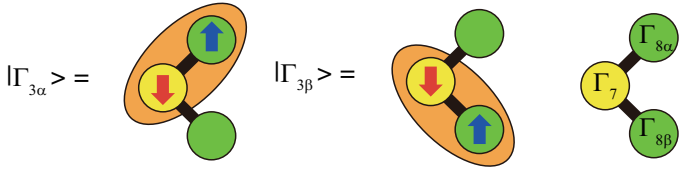}
\caption{
Schematic views for $\Gamma_3$ states composed of two electrons
on $\Gamma_7$ and $\Gamma_{8}$ orbitals of $j=5/2$.
Note that the oval denotes the singlet between $\Gamma_7$ and $\Gamma_8$ states.
The right figure shows the configuration of $\Gamma_7$, $\Gamma_{8\alpha}$,
and $\Gamma_{8\beta}$ orbitals.
}
\end{figure}

We find that the main components of $\Gamma_3$ doublet states are
given by the singlets, $|S_{78\alpha}\rangle$ and $|S_{78\beta} \rangle$
\cite{Kubo}.
It is clearly understood that the $\Gamma_3$ non-Kramers states are non-magnetic
and the quadrupole degrees of freedom are carried out by $\Gamma_8$ orbitals.
Then,  the orbital operators $\bm{T}=(T^x, T^y, T^z)$ are given by
\begin{equation}
\begin{split}
T^z & =\frac{1}{2} (|S_{78\alpha}\rangle \langle S_{78\alpha}|
-|S_{78\alpha}\rangle \langle S_{78\alpha}|),\\
T^{+} &=T^x + i T^y = |S_{78\alpha}\rangle \langle S_{78\beta}|,\\
T^{-} &= T^x-iT^y = |S_{78\beta}\rangle \langle S_{78\alpha}|.
\end{split}
\end{equation}
When we consider the second-order perturbation in terms of the hybridization,
we arrive at the two-channel model with orbital degrees of freedom as
\begin{equation}
  H = \sum_{\bm{k},\tau,\sigma} \varepsilon_{\bm{k}}
    c_{\bm{k}\tau\sigma}^{\dag} c_{\bm{k}\tau\sigma}
   +J (\bm{\tau}_{\uparrow}+\bm{\tau}_{\downarrow}) \cdot \bm{T},
\end{equation}
where $J$ denotes the Kondo exchange coupling and
$\bm{\tau}_{\sigma}=(\tau_{\sigma}^x, \tau_{\sigma}^y, \tau_{\sigma}^z)$
indicates the orbital operator of the conduction electron,
given by
\begin{equation}
\begin{split}
\tau^z_{\sigma} & =\frac{1}{2}\sum_{\bm{k},\bm{k}'}
(c_{\bm{k}\alpha\sigma}^{\dag} c_{\bm{k}'\alpha\sigma}
-c_{\bm{k}\beta\sigma}^{\dag} c_{\bm{k}'\beta\sigma}),\\
\tau^+_{\sigma} & = \tau^x_{\sigma} + i \tau^y_{\sigma}=
\sum_{\bm{k},\bm{k}'}
c_{\bm{k}\alpha\sigma}^{\dag} c_{\bm{k}'\beta\sigma},\\
\tau^-_{\sigma} & = \tau^x_{\sigma} - i \tau^y_{\sigma}
=\sum_{\bm{k},\bm{k}'}
c_{\bm{k}\beta\sigma}^{\dag} c_{\bm{k}'\alpha\sigma}.
\end{split}
\end{equation}

The Hamiltonian $H$ is just the same as the two-channel Kondo model
introduced by Nozi\'eres and Blandin,
when we consider two screening channels for the exchange process
of quadrupole (orbital) degrees of freedom
in a cubic uranium compound with non-Kramers doublet ground state,
as Cox has pointed out.
This model is well known to exhibit the two-channel Kondo effect.

\section{Discussion and Summary}

In this paper, we have confirmed the appearance of
the two-channel Kondo effect in the $\Gamma_3$ non-Kramers doublet state
for the case of $n=2$ by analyzing numerically
the seven orbital impurity Anderson model
hybridized with $\Gamma_8$ conduction bands.
It is true that the two-channel Kondo effect in the $\Gamma_3$ state
has been already discussed for a long time by many researchers,
but we believe that it is meaningful to obtain the two-channel
Kondo effect without considering any assumption
on the CEF excitation at an impurity site.

Concerning the future research development,
we emphasize that it is possible to consider the cases for
all rare-earth ions from Ce$^{3+}$ ($n=1$) to Yb$^{3+}$ ($n=13$).
For the purpose, we assume the hybridization of $\Gamma_8$ conduction electrons
with $j=7/2$ states for the case of $n \ge 7$, whereas we consider
the hybridization between conduction and local $j=5/2$ electrons
for the case of $n<7$.
The results will be shown elsewhere in the future, but here we briefly explain
our recent results for the case of Nd$^{3+}$ ($n=3$) \cite{Hotta3}.

In the seven-orbital impurity Anderson model
hybridized with $\Gamma_8$ conduction electrons,
we have confirmed the two-channel Kondo effect
for the case of $n=3$ with the local $\Gamma_6$ ground state.
To detect the two-channel Kondo effect emerging from Nd ion,
we have proposed to perform the experiments in Nd 1-2-20 compounds.
We expect the appearance of two-channel Kondo effect for other values of
$n$, even for $n>7$ corresponding to heavy rare-earth ion.

In summary, we have shown the two-channel Kondo effect
for the case of $n=2$ with the local $\Gamma_3$ ground state
from the NRG calculations of the seven-orbital impurity Anderson model
hybridized with $\Gamma_8$ conduction electrons.
Since it is possible to change easily the local $f$-electron number,
further studies on the present model are believed to
make significant contributions to the development of new materials
to exhibit the two-channel Kondo effect.

\section*{Acknowledgement}

The author thanks Y. Aoki, K. Hattori, R. Higashinaka, K. Kubo, T. D. Matsuda,
and K. Ueda for useful discussions on heavy-electron systems.
This work was supported by JSPS KAKENHI Grant Number JP16H04017. 
The computation in this work was done using the facilities of the
Supercomputer Center of Institute for Solid State Physics, University of Tokyo.


\appendix

\section{Matrix elements of the effective interactions}

In this Appendix, we explicitly show the equations for the matrix elements of the effective
interactions Eq.~(\ref{effint}).
To save space, we do not separately show ${\tilde I}^{(0)}$ and ${\tilde I}^{(1)}$,
but we exhibit the explicit forms of ${\tilde I}^{aa,aa}$.

Before proceeding to the exhibition of the results, we remark the classification
of the states by using total angular momentum under the cubic CEF potential.
Without the CEF potential, the $j=5/2$ states are specified by $j_z$,
which is the $z$-component of $j$, running between $j_z=-5/2, -3/2, \cdots, 5/2$.
When we include the cubic CEF potential, we obtain $\Gamma_7$ and $\Gamma_8$ states,
but two states of $j_z$ and $j'_z$ with $|j_z-j'_z|=4$ are mixed due to the CEF potential.
Then, the $\Gamma_8$  states are expressed as
\begin{equation}
\begin{split}
f_{a \alpha\uparrow}&=\sqrt{\frac{5}{6}} f_{5/2,-5/2}+\sqrt{\frac{1}{6}} f_{5/2,3/2},\\
f_{a \alpha\downarrow}&=\sqrt{\frac{5}{6}} f_{5/2,5/2}+\sqrt{\frac{1}{6}} f_{5/2,-3/2},\\
f_{a \beta\uparrow}&=f_{5/2,-1/2},\\
f_{a \beta\downarrow}&=f_{5/2,1/2},
\end{split}
\end{equation}
while $\Gamma_7$ states are given by
\begin{equation}
\begin{split}
f_{a \gamma\uparrow}&=\sqrt{\frac{1}{6}} f_{5/2,-5/2}-\sqrt{\frac{5}{6}} f_{5/2,3/2},\\
f_{a \gamma\downarrow}&=\sqrt{\frac{1}{6}} f_{5/2,5/2}-\sqrt{\frac{5}{6}} f_{5/2,-3/2},
\end{split}
\end{equation}
where $f_{j,j_z}$ denotes an annihilation operator of $f$ electron in the basis of
$j$ and $j_z$.

Here we introduce the modified total angular momentum ${\tilde j}$
running among $\pm 3/2$ and $\pm 1/2$,
since the state of $j_z=5/2$ ($-5/2$) belongs to the same group as
that of $j_z=-3/2$ ($3/2$) under the cubic CEF potentials.
Thus, we obtain
${\tilde j}=3/2$ for $\Gamma_{8\alpha\uparrow}$ and $\Gamma_{7\uparrow}$,
${\tilde j}=-3/2$ for $\Gamma_{8\alpha\downarrow}$ and $\Gamma_{7\downarrow}$,
${\tilde j}=1/2$ for $\Gamma_{8\beta\downarrow}$,
and ${\tilde j}=-1/2$ for $\Gamma_{8\beta\uparrow}$.

For the two-electron state
$f^{\dag}_{a \tau_1 \sigma_1} f^{\dag}_{a \tau_2 \sigma_2} |0 \rangle$,
we define the modified total angular momentum ${\tilde J}$
from ${\tilde j}_1 + {\tilde j}_2$.
Then, we classify the two-electron states into four groups,
characterized by ${\tilde J}=0, \pm 1, 2$.

The matrix elements of ${\tilde I}^{(0)}$ calculated from the Coulomb
integrals are expressed by three Racah parameters,
$E_0$, $E_1$, and $E_2$, in Eq.~(\ref{Racahjj}).
Concerning ${\tilde I}^{(1)}$, we derive the matrix elements
from the evaluation of $B_6^0(n,J)({\hat O}_6^0-21{\hat O}_6^4)$
in Eq.~(\ref{StevensH}) with the use of the two-electron states.
In the following, we set $B_6=B_6^0(2,4)$, as defined in Eq.~(\ref{b6}).

For ${\tilde J}=1$, the matrix elements are given by
\begin{equation}
\begin{split}
{\tilde I}^{aa,aa}_{\alpha \uparrow \beta \uparrow, \beta \uparrow \alpha \uparrow}
&=E_0-\frac{13}{3}E_2-24000 B_6, \\
{\tilde I}^{aa,aa}_{\beta \uparrow \gamma \uparrow, \gamma \uparrow  \beta \uparrow}
&=E_0-\frac{5}{3}E_2 + 3480 B_6, \\
{\tilde I}^{aa,aa}_{\gamma \downarrow \alpha \downarrow, \alpha \downarrow  \gamma \downarrow}
&=E_0+5E_2 + 360 B_6, \\
{\tilde I}^{aa,aa}_{\alpha \uparrow \beta \uparrow, \gamma \uparrow  \beta \uparrow}
&=\frac{2\sqrt{5}}{3}E_2 + 1200 \sqrt{5} B_6, \\
{\tilde I}^{aa,aa}_{\alpha \uparrow \beta \uparrow, \alpha \downarrow  \gamma \downarrow}
&=-\frac{2\sqrt{15}}{3}E_2 - 1200 \sqrt{15} B_6, \\
{\tilde I}^{aa,aa}_{\beta \uparrow \gamma \uparrow, \alpha \downarrow  \gamma \downarrow}
&=-\frac{10\sqrt{3}}{3}E_2 + 1560 \sqrt{3} B_6.
\end{split}
\end{equation}

For ${\tilde J}=-1$, the matrix elements are given by
\begin{equation}
\begin{split}
{\tilde I}^{aa,aa}_{\alpha \downarrow \beta \downarrow, \beta \downarrow \alpha \downarrow}
&=E_0-\frac{13}{3}E_2-24000 B_6, \\
{\tilde I}^{aa,aa}_{\beta \downarrow \gamma \downarrow, \gamma \downarrow  \beta \downarrow}
&=E_0-\frac{5}{3}E_2+ 3480 B_6, \\
{\tilde I}^{aa,aa}_{\gamma \uparrow \alpha \uparrow, \alpha \uparrow  \gamma \uparrow}
&=E_0+5E_2+ 360 B_6, \\
{\tilde I}^{aa,aa}_{\alpha \downarrow \beta \downarrow, \gamma \downarrow  \beta \downarrow}
&=\frac{2\sqrt{5}}{3}E_2+ 1200 \sqrt{5} B_6, \\
{\tilde I}^{aa,aa}_{\alpha \downarrow \beta \downarrow, \alpha \uparrow  \gamma \uparrow}
&=-\frac{2\sqrt{15}}{3}E_2 - 1200 \sqrt{15} B_6, \\
{\tilde I}^{aa,aa}_{\beta \downarrow \gamma \downarrow, \alpha \uparrow  \gamma \uparrow}
&=-\frac{10\sqrt{3}}{3}E_2+ 1560 \sqrt{3} B_6.
\end{split}
\end{equation}

For ${\tilde J}=0$, the matrix elements are given by
\begin{equation}
\begin{split}
{\tilde I}^{aa,aa}_{\alpha \uparrow \alpha \downarrow, \alpha \downarrow \alpha \uparrow}
={\tilde I}^{aa,aa}_{\beta \uparrow \beta \downarrow, \beta \downarrow \beta \uparrow}
&= E_0+E_1+2E_2 -7200 B_6, \\
{\tilde I}^{aa,aa}_{\gamma \uparrow \gamma \downarrow, \gamma \downarrow \gamma \uparrow}
&= E_0+E_1-\frac{10}{3}E_2 +67200 B_6, \\
{\tilde I}^{aa,aa}_{\gamma \uparrow \alpha \downarrow, \alpha \downarrow \gamma \uparrow}
={\tilde I}^{aa,aa}_{\alpha \uparrow \gamma \downarrow, \gamma \downarrow \alpha \uparrow}
&= E_0-\frac{10}{3}E_2 +33240 B_6, \\
{\tilde I}^{aa,aa}_{\alpha \uparrow \alpha \downarrow, \beta \downarrow \beta \uparrow}
&= E_1-\frac{11}{3}E_2 -26400 B_6, \\
{\tilde I}^{aa,aa}_{\alpha \uparrow \alpha \downarrow, \gamma \downarrow \gamma \uparrow}
={\tilde I}^{aa,aa}_{\beta \uparrow \beta \downarrow, \gamma \downarrow \gamma \uparrow}
&= E_1+\frac{5}{3}E_2 +33600 B_6, \\
{\tilde I}^{aa,aa}_{\alpha \uparrow \alpha \downarrow, \alpha \downarrow \gamma \uparrow}
={\tilde I}^{aa,aa}_{\alpha \uparrow \alpha \downarrow, \gamma \downarrow \alpha \uparrow}
&= \frac{4\sqrt{5}}{3}E_2 -7680 \sqrt{5} B_6, \\
{\tilde I}^{aa,aa}_{\beta \uparrow \beta \downarrow, \alpha \downarrow \gamma \uparrow}
={\tilde I}^{aa,aa}_{\beta \uparrow \beta \downarrow, \gamma \downarrow \alpha \uparrow}
&= -\frac{4\sqrt{5}}{3}E_2 +7680 \sqrt{5}B_6, \\
{\tilde I}^{aa,aa}_{\gamma \uparrow \alpha \downarrow, \gamma \downarrow \alpha \uparrow}
&= \frac{5}{3}E_2 +28200 B_6.
\end{split}
\end{equation}

For ${\tilde J}=2$, the matrix elements are given by
\begin{equation}
\begin{split}
{\tilde I}^{aa,aa}_{\alpha \uparrow \beta \downarrow, \beta \downarrow \alpha \uparrow}
={\tilde I}^{aa,aa}_{\beta \uparrow \alpha \downarrow, \beta \downarrow \beta \uparrow}
&= E_0+\frac{2}{3}E_2 -2400 B_6, \\
{\tilde I}^{aa,aa}_{\beta \uparrow \gamma \downarrow, \gamma \downarrow \beta \uparrow}
={\tilde I}^{aa,aa}_{\gamma \uparrow \beta \downarrow, \beta \downarrow \gamma \uparrow}
&= E_0-\frac{10}{3}E_2 +30120 B_6, \\
{\tilde I}^{aa,aa}_{\alpha \uparrow \beta \downarrow, \alpha \downarrow \beta \uparrow}
&= 5E_2 +21600 B_6, \\
{\tilde I}^{aa,aa}_{\beta \uparrow \gamma \downarrow, \beta \downarrow \gamma \uparrow}
&= -5E_2 +31320 B_6, \\
{\tilde I}^{aa,aa}_{\alpha \uparrow \beta \downarrow, \gamma \downarrow \beta \uparrow}
={\tilde I}^{aa,aa}_{\beta \uparrow \alpha \downarrow, \beta \downarrow \gamma \uparrow}
&= -2\sqrt{5}E_2 +6480\sqrt{5} B_6, \\
{\tilde I}^{aa,aa}_{\alpha \uparrow \beta \downarrow, \beta \downarrow \gamma \uparrow}
={\tilde I}^{aa,aa}_{\beta \uparrow \gamma \downarrow, \alpha \downarrow \beta \uparrow}
&= -\frac{2\sqrt{5}}{3}E_2 +8880\sqrt{5} B_6.
\end{split}
\end{equation}
Note the relation of
${\tilde I}^{aa,aa}_{\tau_4 \sigma_4 \tau_3 \sigma_3,\tau_2 \sigma_2 \tau_1 \sigma_1}
={\tilde I}^{aa,aa}_{\tau_1 \sigma_1 \tau_2 \sigma_2,\tau_3 \sigma_3 \tau_4 \sigma_4}$.

To check the above matrix elements,
we diagonalize the Coulomb matrix for each ${\tilde J}$
and obtain 15 eigenenergies in total.
Among them, the nonet of $J=4$ is split into four groups as
$\Gamma_1$ singlet with $E_0-5E_2-100800 B_6$,
$\Gamma_3$ doublet with $E_0-5E_2 + 80640 B_6$,
$\Gamma_4$ triplet with $E_0-5E_2 + 5040 B_6$,
and
$\Gamma_5$ triplet with $E_0-5E_2 -25200 B_6$.
Note that they are equal to the CEF energies of $f^2$ states
for the case of $B_4^0=0$.
The quintet of $J=2$ and the singlet of $J=0$ are not influenced by
the $B_6^0$ term and their energies are determined only by
the Coulomb interactions, leading to $E_0 + 9E_2$ for $J=2$
and $E_0 + 3E_1$ for $J=0$, respectively.
We note that the quintet of $J=2$ is split into two groups of $\Gamma_3$
doublet and $\Gamma_5$ triplet, when we consider the $B_4^0$ term
of the one-electron potential.

Finally, it is instructive to pick up the interactions between $\Gamma_8$ states,
leading to a two-orbital local Hamiltonian, given by
\begin{equation}
\begin{split}
H &= U \sum_{\tau} \rho_{\tau\uparrow} \rho_{\tau\downarrow}
+ U' \rho_{\alpha} \rho_{\beta}
+ J \sum_{\sigma,\sigma'}
f_{\alpha\sigma}^{\dag} f_{\beta\sigma'}^{\dag}
f_{\alpha\sigma'} f_{\beta\sigma} \\
&+J' (f_{\alpha\uparrow}^{\dag} f_{\alpha \downarrow}^{\dag}
f_{\beta\downarrow} f_{\beta\uparrow} + {\rm h.c.}),
\end{split}
\end{equation}
where $\rho_{\tau\sigma}=f^{\dag}_{\tau\sigma}f_{\tau\sigma}$
and
$\rho_{\tau}=\rho_{\tau\uparrow}+\rho_{\tau\downarrow}$.
The coupling constants $U$, $U'$, $J$, and $J'$ denote
intra-orbital, inter-orbital, exchange, and pair-hopping interactions,
respectively, expressed by $E_0$, $E_1$, $E_2$, and $B_6$ as
\begin{equation}
\begin{split}
U &= E_0+E_1+2E_2-7200 B_6,\\
U' &= E_0+\frac{2}{3}E_2-2400 B_6,\\
J &= 5E_2+21600 B_6,\\
J'&=E_1-\frac{11}{3}E_2-26400 B_6.
\end{split}
\end{equation}
Note the relation of $U=U'+J+J'$, ensuring the rotational invariance
in the orbital space for the interaction part.
We also note that the relation of $U=U'+J+J'$ holds even for $B_6=0$.
On the other hand, the relation of $J=J'$ does not hold in the present case
in sharp contrast to the $d$-electron case.
It is quite natural, since this relation is due to the reality of
the wavefunction and in the $j$-$j$ coupling scheme,
the wavefunction is complex.
When we effectively include the effect of the sixth-order CEF potentials
in the $\Gamma_8$ model, it is reasonable to consider the situation
of $J<0$, leading to the stabilization of the $\Gamma_3$ state.

\bigskip


\begin{thebibliography}{99}


\bibitem{Nozieres}
Ph. Nozi\'eres and A. Blandin,
J. Physique {\bf 41}, 193 (1980).

\bibitem{Jones1}
B. A. Jones and C. M. Varma,
Phys. Rev. Lett. {\bf 58}, 843 (1987).

\bibitem{Jones2}
B. A. Jones, C. M. Varma, and J. W. Wilkins,
Phys. Rev. Lett. {\bf 61}, 125 (1988).

\bibitem{Cox1}
D. L. Cox, Phys. Rev. Lett. {\bf 59}, 1240 (1987).

\bibitem{Cox2}
D. L. Cox and A. Zawadowski,
{\it Exotic Kondo Effects in Metals}
(Taylor \& Francis, London, 1999), p. 24.

\bibitem{review}
See, for instance, 
T. Onimaru and H. Kusunose, J. Phys. Soc. Jpn. {\bf 85}, 082002 (2016)
and references therein.

\bibitem{Slater}
J. C. Slater,
{\it Quantum Theory of Atomic Structure} (McGraw-Hill, New York, 1960).

\bibitem{Hutchings}
M. T. Hutchings,
Solid State Phys. {\bf 16}, 227 (1964).

\bibitem{Carnall}
W. T. Carnall, P. R. Fields, and K. Rajnak,
J. Chem. Phys. {\bf 49}, 4424 (1968).

\bibitem{NRG1}
K. G. Wilson, Rev. Mod. Phys. {\bf 47}, 773 (1975).

\bibitem{NRG2}
H. R. Krishna-murthy, J. W. Wilkins, and K. G. Wilson,
Phys. Rev. B {\bf 21}, 1003 (1980).

\bibitem{Fabrizio1}
M. Fabrizio, A. F. Ho, L. D. Leo, and G. E. Santoro,
Phys. Rev. Lett. {\bf 91}, 246402 (2003).

\bibitem{Fabrizio2}
L. D. Leo and M. Fabrizio,
Phys. Rev. B {\bf 69}, 245114 (2004).

\bibitem{Mitchell}
A. K. Mitchell and E. Sela,
Phys. Rev. B {\bf 85}, 235127 (2012).

\bibitem{Miyake1}
S. Yotsuhashi,  K. Miyake, and H. Kusunose,
J. Phys. Soc. Jpn. {\bf 71}, 389 (2002).

\bibitem{Miyake2}
S. Nishiyama, H. Matsuura, and K. Miyake,
J. Phys. Soc. Jpn. {\bf 79}, 104711 (2010).

\bibitem{Miyake3}
S. Nishiyama and K. Miyake,
J. Phys. Soc. Jpn. {\bf 80}, 124706 (2011).

\bibitem{Hotta1}
T. Hotta and K. Ueda, Phys. Rev. B {\bf 67}, 104518 (2003).

\bibitem{Hotta2}
T. Hotta and H. Harima, J. Phys. Soc. Jpn. {\bf 75}, 124711 (2006).

\bibitem{Hattori}
K. Hattori,  T. Nomoto, T. Hotta, and H. Ikeda, preprint.

\bibitem{Stevens}
K. W. H. Stevens, Proc. Phys. Soc. A{\bf 65} (1952) 209.

\bibitem{Kubo}
K. Kubo and T. Hotta, Phys. Rev. B {\bf 95}, 054425 (2017).

\bibitem{Hotta3}
T. Hotta, J. Phys. Soc. Jpn. {\bf 86}, 083704 (2017).

\end{thebibliography}


\end{document}